\documentclass[12pt,a4paper]{article}
\usepackage[cp1251]{inputenc}
\usepackage{epsfig,ifthen,amsmath,amssymb,amsfonts}
\usepackage[english]{babel}
\usepackage[T2A]{fontenc}
\usepackage[colorlinks]{hyperref}
\usepackage{verbatim}
\usepackage{indentfirst}

\begin{document}

\title{Quantum correlations of twophoton polarization states in the
parametric down-conversion process}
\author{Ivan S. Dotsenko
\thanks{E-mail: ivando@ukr.net} \ \ \ \ \ Volodymyr G. Voronov
\thanks{E-mail: ktpist@ukr.net} \\
\it \small Faculty of Physics, Taras Shevchenko Kyiv National University,\\
\it \small 6 Academician Glushkov Ave., 03680 Kyiv, Ukraine}
\date{}
\maketitle

\begin{abstract}
We consider correlation properties of twophoton polarization states
in the parametric down-conversion process. In our description of
polarization states we take into account the simultaneous presence
of colored and white noise in the density matrix. Within the
considered model we study the dependence of the von Neumann entropy
on the noise amount in the system and derive the separability
condition for the density matrix of twophoton polarization state,
using Perec-Horodecki criterion and majorization criterion. Then the
dependence of the Bell operator (in CHSH form) on noise is studied.
As a result, we give a condition for determining the presence of
quantum correlation states in experimental measurements of the Bell
operator. Finally, we compare our calculations with experimental
data~\cite{Bovino} and give a noise amount estimation in the photon
polarization state considered there.
\end{abstract}

\section{Introduction}
In 1982 Aspect's group (Alain Aspect et. al.~\cite{Aspect})
performed  a verification experiment for possible violation of
Bell's inequalities in Clauser-Horne-Shimony-Holt (CHSH)
form~\cite{CHSH}, where a correlation measurement of twophoton
polarization states was provided. Experimental data gave Bell's
inequality violation by five standard deviations. Measurement
results corresponded well with predictions of quantum mechanics.
Numerous later experiments showed that their results are in
agreement with the quantum mechanical description of nature.

Thus, specific quantum correlations obtained the status of reality,
and entangled states, which provide such correlations, became an
object of intensive research. It turned out that entanglement can
play in essence the role of a new resource in such scientific areas
as quantum cryptography, quantum teleportation, quantum
communication and quantum computation. This became a great stimulus
for researching methods of creating, accumulating, distributing and
broadcasting this resource.

One of the most important questions in the considered topic concerns
methods of identifying the presence of entanglement in one or
another realistic quantum mechanical state. Since entangled states
violate Bell's inequalities, the violation of Bell's inequalities
can be a basic tool to detect entanglement. In realistic
applications pure entangled states become mixed states due to
different types of noise. Thus a question about robustness of Bell's
inequalities violation against the noise arises. In other words, one
wants to know, under what proportion of an entangled state and noise
in a realistic mixed state the presence of entanglement can be
discovered.

The most reliable source of two-party entanglement are
polarization-entangled photons created by the parametric
down-conversion process (PDC)~\cite{PDC}.

\section{Noise-present entanglement detection}
In 2006 a paper by Bovino (Fabio A. Bovino et al.~\cite{Bovino})
appeared. It concerned the experimental verification of the CHSH
inequality robustness against colored noise. A crystal (beta-barium
borate) was irradiated by a laser, working in pulsed mode, and in
the PDC process photon pairs in polarization-correlated states were
created. These states correspond to the following polarization
density matrix:
\begin{equation}
\label{rho_colored} \hat{\rho} =
p|\Phi^+\rangle\langle\Phi^+|+\frac{1-p}{2} (|00\rangle\langle
00|+|11\rangle\langle 11|),
\end{equation}
where
$|\Phi^+\rangle=\frac{1}{\sqrt{2}}\left(|00\rangle+|11\rangle\right)$
is  one of the four entangled Bell's states. State $|1\rangle$
corresponds to ordinary polarization and state $|0\rangle$
correponds to extraordinary ray polarization in the uniaxial
crystal.

In the current paper theoretical analysis for robustness of Bell's
inequality (in CHSH form) violation with simultaneous presence of
colored and white noise is performed. The density matrix for the
twophoton polarization state in such a generalized model can be
expressed in the form:
\begin{equation}
\label{rho_CW} \hat{\rho}_{CW} =
p|\Phi^+\rangle\langle\Phi^+|+\frac{r}{2}
(|00\rangle\langle00|+|11\rangle\langle11|)+\frac{1-(p+r)}{4}\hat{I}.
\end{equation}
Varying the parameter $p$ in the range from $0$ to $1$, one can
change the pure state $|\Phi^+\rangle$ fraction in \eqref{rho_CW},
and changing $r$ from $0$ to $(1-p)$, with the value of $p$ fixed,
one can adjust relative colored and white noise fractions.

For $r=0$ we have the particular case of colored noise absence:
\begin{equation}
\label{rho_white}
\hat{\rho}_W=p|\Phi^+\rangle\langle\Phi^+|+\frac{1-p}{4}\hat{I},
\end{equation}
where $\hat{I}$ is the $4\times4$ identity matrix. These states are
called Werner states~\cite{Werner}. And for $r=1-p$ we have
\eqref{rho_colored}, which is the case of white noise absence.

%===PART === FOR === TRANSLATION=======
%===translate it!======================

Examine first the general structure of the density matrix
\eqref{rho_CW}. In the basic states representation $\{ |00\rangle$,
$|01\rangle$, $|10\rangle$, $|11\rangle \}$ the density matrix looks
as following:
\begin{equation}
\label{matrix_rho_CW} \rho_{CW}= \left(
  \begin{array}{cccc}
    \frac{1}{4}(p+r+1) & 0 & 0 & \frac{1}{2}p \\
    0 & \frac{1}{4}(1-p-r) & 0 & 0 \\
    0 & 0 & \frac{1}{4}(1-p-r) & 0 \\
    \frac{1}{2}p & 0 & 0 & \frac{1}{4}(p+r+1) \\
  \end{array}
\right),
\end{equation}
while in the Bell's states representation $\{ \
|\Phi^+\rangle=\frac{1}{\sqrt{2}}(|00\rangle+|11\rangle)$, \ \
$|\Phi^-\rangle=\frac{1}{\sqrt{2}}(|00\rangle-|11\rangle)$, \ \
$|\Psi^+\rangle=\frac{1}{\sqrt{2}}(|01\rangle+|10\rangle)$, \ \
$|\Psi^-\rangle=\frac{1}{\sqrt{2}}(|01\rangle-|10\rangle) \ \}$ the
density matrix is diagonal:
\begin{equation}
\label{matrix_rho_diag} \rho^{diag}_{CW}= \left(
  \begin{array}{cccc}
    \frac{1}{4}(1+3p+r) & 0 & 0 & 0 \\
    0 & \frac{1}{4}(1-p+r) & 0 & 0 \\
    0 & 0 & \frac{1}{4}(1-p-r) & 0 \\
    0 & 0 & 0 & \frac{1}{4}(1-p-r) \\
  \end{array}
\right).
\end{equation}
Numbers $\lambda_i$, which are on the diagonal, are the eigenvalues
of the density matrix \eqref{matrix_rho_CW}.

Thus, \eqref{rho_CW} can be represented by means of the projector
operators on the Bell's states:
\begin{equation}
\begin{split}
\label{projectors} \rho_{CW}=
\frac{1}{4}(1+3p+r)|\Phi^+\rangle\langle\Phi^+| +
\frac{1}{4}(1-p+r)|\Phi^-\rangle\langle\Phi^-| +{}\\
+\frac{1}{4}(1-p-r)|\Psi^+\rangle\langle\Psi^+| +
\frac{1}{4}(1-p-r)|\Psi^-\rangle\langle\Psi^-|.
\end{split}
\end{equation}

For $p=0$ and $r=0$ all basic states go into \eqref{projectors} with
equal weight coefficients $W=\frac{1}{4}$, i.e. the density operator
is proportional to unit operator, and for $p=1$, $r=0$: $
\rho_{CW}=|\Phi^+\rangle\langle\Phi^+|$ we have a pure state.

\begin{figure}[tbh]
\centering \epsfig{file=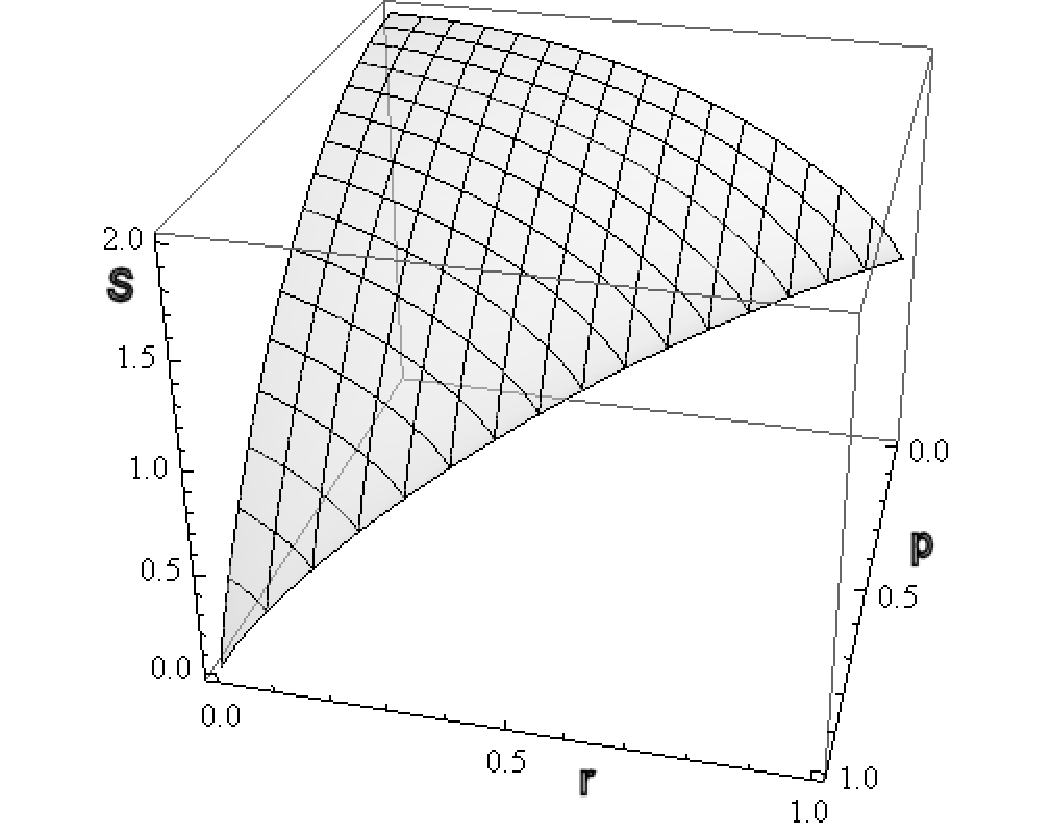, height=6.5cm} \caption{The
surface of the von Neumann entropy values of the density matrix
\eqref{rho_CW} depending on the values of $p$ and $r$
parameters.}\label{fig:3d_15_11}
\end{figure}

In the Fig. \ref{fig:3d_15_11} the von Neumann entropy dependence as a
($p$,$r$)-parameter function is represented:
$S(\rho_{CW})=S(p,r)=-Tr(\rho\log\rho)=-\sum_i\lambda_i\log\lambda_i$.

For $p=1$, $r=0$ the von Neumann entropy is zero, and for $p=0$,
$r=0$ it reaches its maximal value $S=2$.

The matrix obtained from \eqref{matrix_rho_CW} by partial transpose
of states of the first subsystem is the following:
\begin{equation}
\label{matrix_rho} \rho^{T_A}= \left(
  \begin{array}{cccc}
    \frac{1}{4}(p+r+1) & 0 & 0 & 0 \\
    0 & \frac{1}{4}(1-p-r) & \frac{1}{2}p & 0 \\
    0 & \frac{1}{2}p & \frac{1}{4}(1-p-r) & 0 \\
    0 & 0 & 0 & \frac{1}{4}(p+r+1) \\
  \end{array}
\right),
\end{equation}
and after the diagonalization:
\begin{equation}
\label{matrix_rho_trans} \rho^{T_A}_{diag}= \left(
  \begin{array}{cccc}
    \frac{1}{4}(p+r+1) & 0 & 0 & 0 \\
    0 & \frac{1}{4}(1+p+r) & 0 & 0 \\
    0 & 0 & \frac{1}{4}(1+p-r) & 0 \\
    0 & 0 & 0 & \frac{1}{4}(1-3p-r) \\
  \end{array}
\right).
\end{equation}
Here eigenvalues $\lambda^T_1=\frac{1}{4}(1+p+r), \ \
\lambda^T_2=\frac{1}{4}(1+p+r), \ \ \lambda^T_3=\frac{1}{4}(1+p-r),
\ \ \lambda^T_4=\frac{1}{4}(1-3p-r)$ are given in a way, that they
satisfy the inequality:
\begin{equation}
\lambda^T_1 \geq \lambda^T_2 \geq \lambda^T_3 \geq \lambda^T_4.
\end{equation}

Since $\rho_{CW}$ in \eqref{rho_CW} is valid only for $0\leq p \leq
1$, $0\leq p+r \leq1$, then \ $\lambda^T_1$, $\lambda^T_2$,
$\lambda^T_3$ are nonnegative for any valid values of $p$ and $r$,
while $\lambda^T_4$ is negative for $3p+r>1$. According to the
Perec-Horodecki criterion~\cite{Perec, Horod_2} for systems, which
consist of two subsystems with the dimensions $m\times n \leq 6$,
where $m$ and $n$ are dimensions of first and second subsystem,
respectively, the necessary and sufficient condition for the state
separability is the condition of non-negativity of all eigenvalues
of the density matrix $\rho^{T_A}$.

In the case, considered in present paper, the state \eqref{rho_CW}
is separable, and thus unentangled, under the following condition:
\begin{equation}
\left\{
  \begin{array}{ll}
    3p+r<1, & \hbox{ } \\
    0<p<1. & \hbox{ }
  \end{array}
\right.
\end{equation}

%=====================================

\begin{figure}[tbh]
\centering \epsfig{file=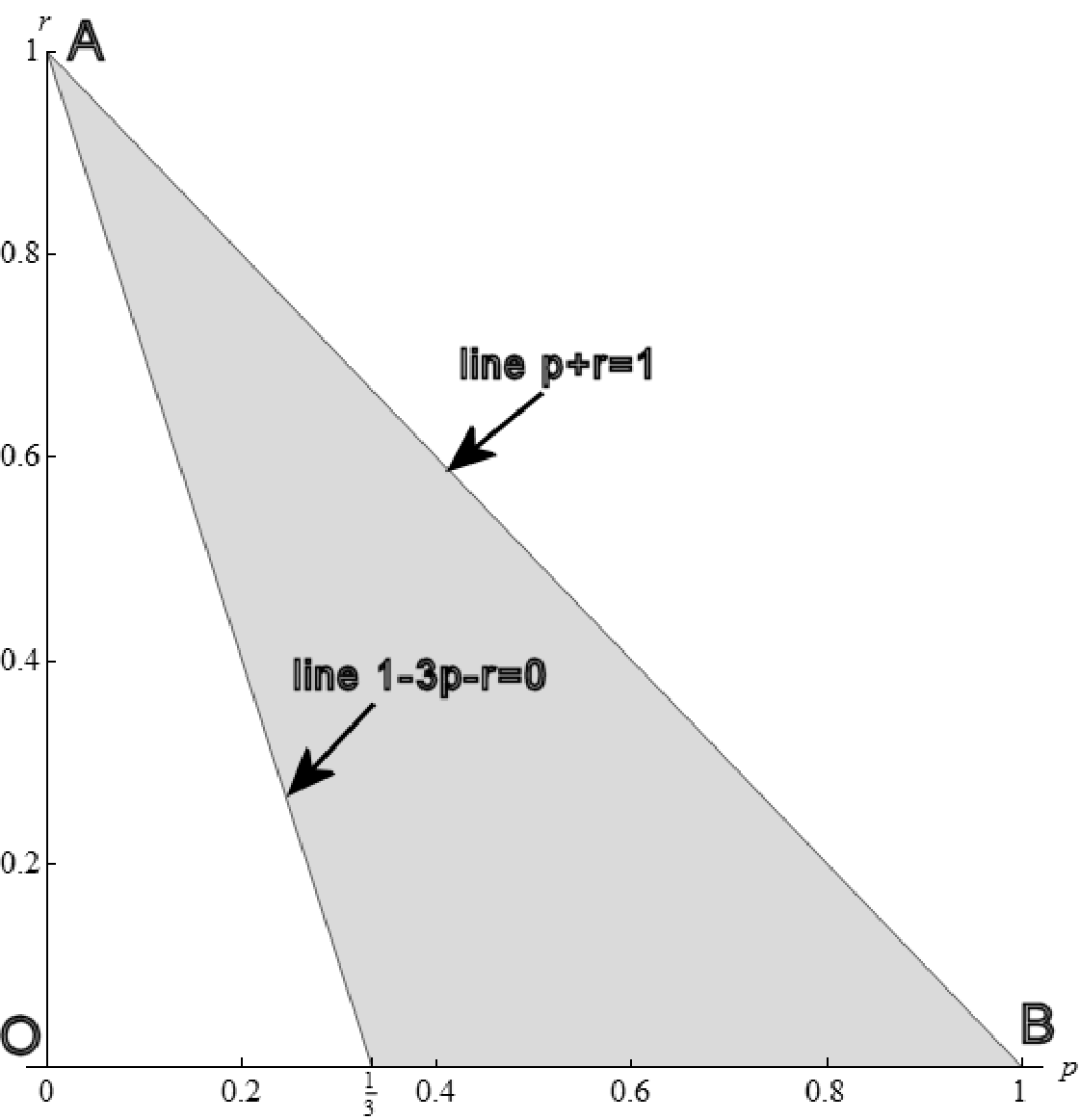, height=8.6cm}
\caption{The filled area corresponds to the set of values of $p$ and
$r$ parameters, for which the state \eqref{rho_CW} is
inseparable.}\label{fig:trikutnik_eng}
\end{figure}

In the Fig. \ref{fig:trikutnik_eng} the filled area in the $OAB$
triangle corresponds to inseparable (entangled) states. For a fixed
value $p<\frac{1}{3}$ a separable state can become inseparable, if
one increases the colored noise fraction while reducing the white
noise fraction (by increasing the value of the $r$ parameter). For
the Werner state ($r=0$) we obtain the well-known
result\cite{Werner}: the state is separable for $p<\frac{1}{3}$.

For $p+r=1$ (white noise absence) one obtains
$\lambda_4^T=-\frac{1}{2}p <0$, which corresponds well to the known
statement~\cite{Cabello} that under presence of some colored noise
fraction and simultaneous absence of white noise the state
\eqref{rho_CW} is remaining entangled (inseparable).

The reduced density matrix of the first and the second subsystem in
the state \eqref{rho_CW} is proportional to the unit matrix:
$\rho^A=\frac{1}{2}\hat{I}$, $\rho^B=\frac{1}{2}\hat{I}$ and
independent from $p$ and $r$. Thus, measurement of the polarization
state of a single photon in \eqref{rho_CW} in any orthogonal basis
gives the same result.

We now apply the majorization criterion~\cite{Nielsen} to the state
\eqref{rho_CW}.

According to the criterion, if a density matrix $\rho$ is separable,
then the following condition is satisfied:
\begin{equation}
\lambda_{\rho}^{\downarrow} \prec \lambda_{\rho_A}^{\downarrow} \ \
\ \text{and} \ \ \ \lambda_{\rho}^{\downarrow} \prec
\lambda_{\rho_B}^{\downarrow},
\end{equation}
where $\lambda_{\rho}^{\downarrow}$ denotes the vector, whose
components are the eigenvalues of the matrix $\rho$, put in the
nonincreasing order, and one can say, that the vector
$x^{\downarrow}$ is majorized by the vector $y^{\downarrow}$, if
$\sum^k_{i=1}x^{\downarrow}_i \leq \sum^k_{i=1}y^{\downarrow}_i$,
$k=1,2,3,\ldots,d$, where $d$ is the dimension of the Hilbert space
of states and equality is achieved if and only if $k=d$.

In our case
$$ x^{\downarrow}=
\left(
                                   \begin{array}{c}
                                     \lambda_1 \\
                                     \lambda_2 \\
                                     \lambda_3 \\
                                     \lambda_4 \\
                                   \end{array}
                                 \right), \ \ \ \ \ \
y^{\downarrow}= \left(
                                   \begin{array}{c}
                                     \frac{1}{2} \\
                                     \frac{1}{2} \\
                                     0 \\
                                     0 \\
                                   \end{array}
                                 \right),
$$
where $\lambda_1=\frac{1}{4}(1+3p+r)$, \
$\lambda_2=\frac{1}{4}(1-p+r)$, \ $\lambda_3=\frac{1}{4}(1-p-r)$, \
$\lambda_4=\frac{1}{4}(1-p-r)$.

Then according to the majorization criterion:
$$ \left\{
  \begin{array}{ll}
    \lambda_1<\frac{1}{2} & \hbox{ } \\
    \lambda_1+\lambda_2 <\frac{1}{2}+\frac{1}{2} & \hbox{ } \\
    \lambda_1+\lambda_2+\lambda_3<\frac{1}{2}+\frac{1}{2}+0  & \hbox{ } \\
    \lambda_1+\lambda_2+\lambda_3+\lambda_4=\frac{1}{2}+\frac{1}{2}+0+0 & \hbox{ }
  \end{array}
\right. \Rightarrow \left\{
  \begin{array}{ll}
    3p+r<1 & \hbox{ } \\
    p+r<1 & \hbox{ } \\
    p+r<1 & \hbox{ } \\
    1=1 & \hbox{ }
  \end{array}
\right. .$$ The second and the third inequality are always
satisfied, if the first inequality is satisfied. Therefore, the
state \eqref{rho_CW} is separable, if the condition $3p+r<1$ is
satisfied, which coincides with the condition, obtained from the
Perec-Horodecki criterion.

%========================================
%===END OF THE PART =====================
%========================================
%===as old one===========================
Consider now, under what conditions the state \eqref{rho_CW}
violates the Bell's inequality in the CHSH form
\begin{equation}
\label{CHSH} |\beta|\leq2,
\end{equation}
where
\begin{equation}
\label{Bell_operator} \ \beta=-\langle A_0B_0\rangle-\langle
A_0B_1\rangle-\langle A_1B_0\rangle+ \langle A_1B_1\rangle
\end{equation}
is called the Bell operator.

For maximal Bell's inequality \eqref{CHSH} violation  analysis,
separately in states with white \eqref{rho_white} and colored
\eqref{rho_colored} noise, in a paper by Cabello (Adan Cabello at
al.~\cite{Cabello}) the following onequbit observables were taken:
\begin{equation}
\label{A_B} \left\{
  \begin{array}{ll}
    A_0=\sigma_z, & \hbox{ } \\
    A_1=\cos(\theta)\sigma_z+sin(\theta)\sigma_x, & \hbox{ } \\
    B_0=\cos(\phi)\sigma_z+sin(\phi)\sigma_x, & \hbox{ } \\
    B_1=\cos(\phi-\theta)\sigma_z+sin(\phi-\theta)\sigma_x. & \hbox{ }
  \end{array}
\right.
\end{equation}

The $\theta$ and $\phi$ parameters in \eqref{A_B} determine the
orientation of analyzers in experimental devices, $\sigma_x$ and
$\sigma_z$ are the usual Pauli matrices. Computations showed, that
for the Werner state \eqref{rho_white} the maximal value of $\beta$
as a $p$-parameter function is the following:
\begin{equation}
\beta_{max}(p)=2\sqrt{2}p
\end{equation}
and for all values of $p$ the maximal value $\beta$ is obtained by
$\theta=\frac{\pi}{2}$, $\phi=\frac{\pi}{4}$.

Thus, Bell's inequality \eqref{CHSH} is violated only  for
$p>1/\sqrt{2}\approx0.707$. This implies, that in the case, when the
entangled state $|\Phi^+\rangle$ is distorted only by white noise,
entanglement presence can be detected if noise proportion is less
then $\thicksim29\%$.

In the presence of colored noise \eqref{rho_colored} the maximal
value of $\beta$ for different values of $p$ is achieved at
different values of angles $\theta$ and $\phi$. The most interesting
fact is that the state \eqref{rho_colored} violates the CHSH
inequality for all values $0<p\leq1$. Thus, Bell's inequality
violation is extremely robust against colored noise.

%==========================
In the state \eqref{rho_CW} the quantity $\beta$, which responds to
onequbit observables \eqref{A_B} is a four-parameter function:
\begin{equation}
\begin{split}
\label{beta_CW}
\beta_{CW}(p,r,\theta,\phi)=cos(\phi)[(2p+r)(sin^2(\theta)+cos(\theta))+{}\\
+rcos(\theta)]-sin(\phi)(2p+r)[cos(\theta)-1]sin(\theta).
\end{split}
\end{equation}
In the colored noise absence $(r=0)$ we have:
\begin{equation}
\beta_W(p,\theta,\phi)=2p\{cos(\phi)[sin^2(\theta)+cos(\theta)]-sin(\phi)[cos(\theta)-1]sin(\theta)\},
\end{equation}
and in the white noise absence $(r=1-p)$:
\begin{equation}
\beta_C(p,\theta,\phi)=cos(\phi)[(1+p)sin^2(\theta)+2cos(\theta)]-sin(\phi)(1+p)[cos(\theta)-1]sin(\theta).
\end{equation}

For fixed values of the $p$ and $r$ parameters the expression
\eqref{beta_CW} is a function of $\theta$ and $\phi$. Solving the
extremum problem for the two-variable function, one can find the
maximal values $\beta^{max}_{CW}(p,r)$, as well as the angles
$\theta$ and $\phi$, that provide the maximal $\beta_{CW}(p,r)$.

\begin{figure}[tbh!]
\centering \epsfig{file=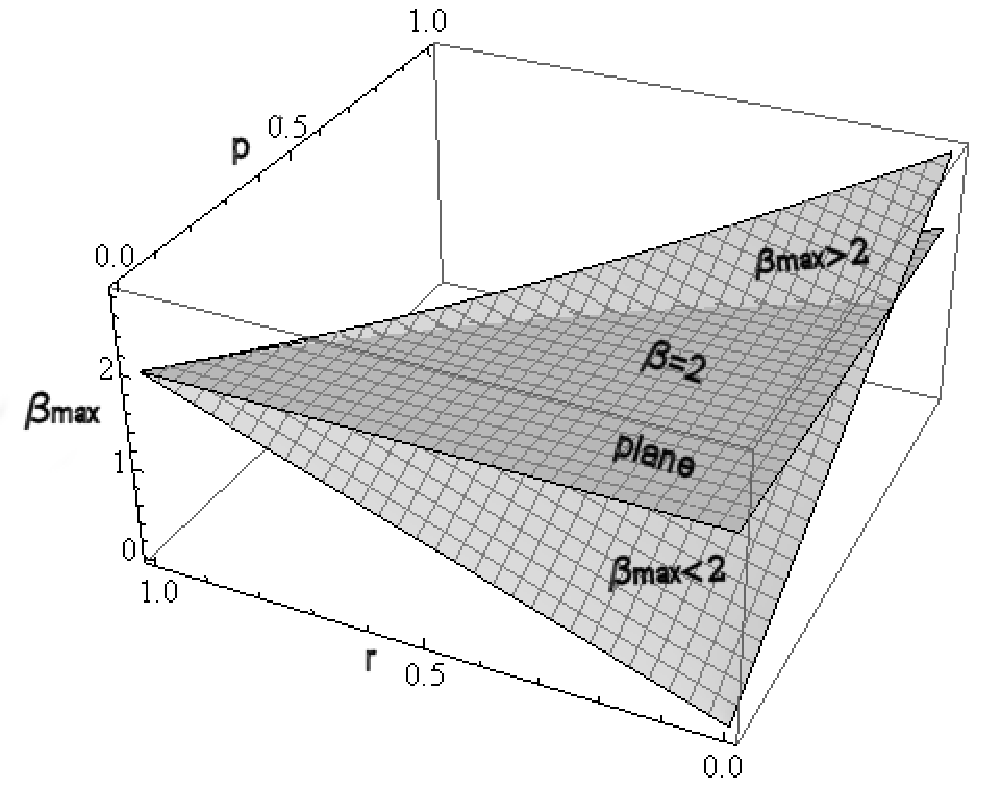, height=6cm} \caption{3D Plot for the
maximal Bell operator values, and the $\beta_{CW}=2$ plane, that
corresponds to the classical bound.}\label{fig:3d}
\end{figure}

%==============================================================
In the Fig. \ref{fig:3d} the shaded surface graphically displays the
 $\beta^{max}_{CW}(p,r)$ as a function of two variables $p$ and $r$.
For comparison the plane $\beta=2$, which is the boundary value of
Bell's inequality, is also represented in the figure. The surface
patch above the plane $\beta=2$ is the CHSH inequality violation
area.

\begin{figure}[tbh!]
\centering \epsfig{file=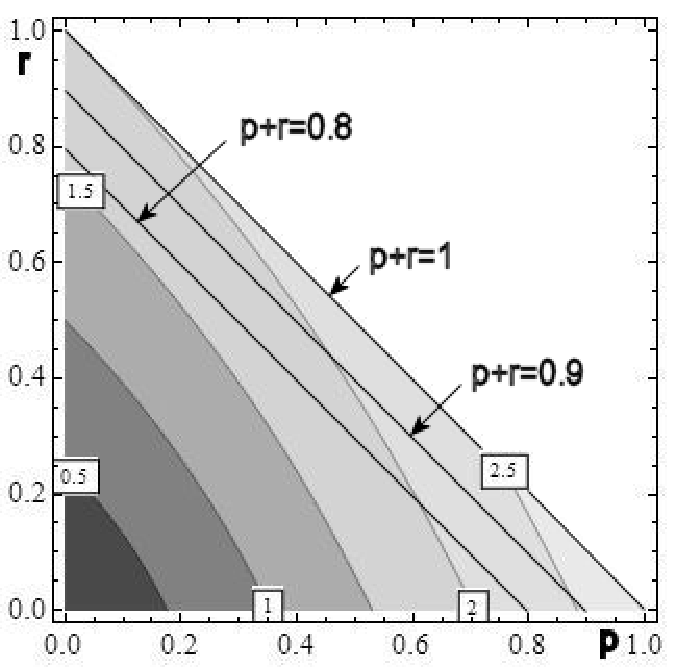, height=8cm}
\caption{Contour plot for $\beta^{max}(p,r)=const$ - maximal Bell
operator values on coordinate plane
$(p,r)$.}\label{fig:contour_new2}
\end{figure}

In the Fig. \ref{fig:contour_new2} projections of the traces
$\beta=const$ on the $(p,r)$-plane with the surface
$\beta^{max}_{CW}(p,r)$ are represented. From the figure one can see
that the straight line $p+r=1$ (white noise absence) fully lies in
the $\beta^{max}>2$ area, which corresponds to the above conclusion,
that Bell's inequality violation is robust against colored noise.
For $r=0$ (colored noise absence) Bell's inequality is violated only
for $p>1/\sqrt{2}$. For any fixed $p$ (pure entangled state weight
factor) the value of $\beta^{max}_{CW}$ decreases with the
increasing white noise fraction. Thus, as expected, adding some
amount of white noise to colored one can reach better agreement of
theoretically computed $\beta^{max}$ values with experimental ones.
Bell's inequality violation is unsteady under the increasing of
white noise fraction for a fixed total amount (white and colored) of
noise.

\begin{figure}[tbh!]
\centering \epsfig{file=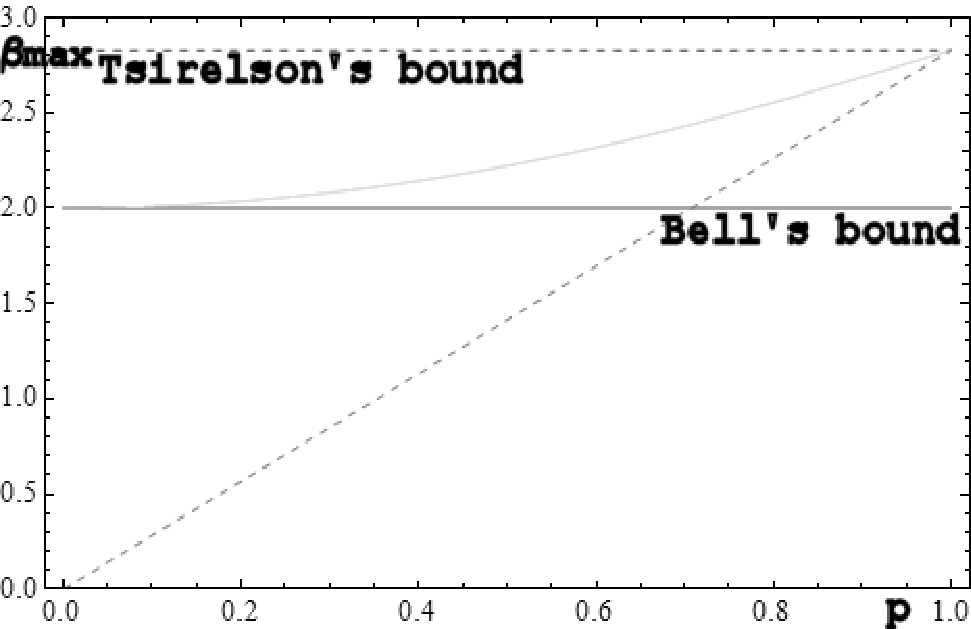, height=6cm} \caption{Maximal Bell
operator values in the case $r=1-p$ is white noise absence(top
curve) and $r=0$ is colored noise absence (bottom dashed straight
line). Classical bound is $2$. Tsirelson's bound~\cite{Tsirelson} is
$2\sqrt{2}=2.83$.}\label{fig:2d}
\end{figure}

%===================
In the Fig. \ref{fig:2d} the $\beta^{max}_{CW}(p,r)$ dependence on $p$ in two
boundary cases is given: $r+p=1$ is white noise absence (top curve)
and $r=0$ is colored noise absence (bottom dashed straight line).
The boundary case dependencies of $\beta^{max}$ on $p$ and $r$
coincide with the ones from the work~\cite{Bovino}.

\begin{figure}[tbh!]
\centering \epsfig{file=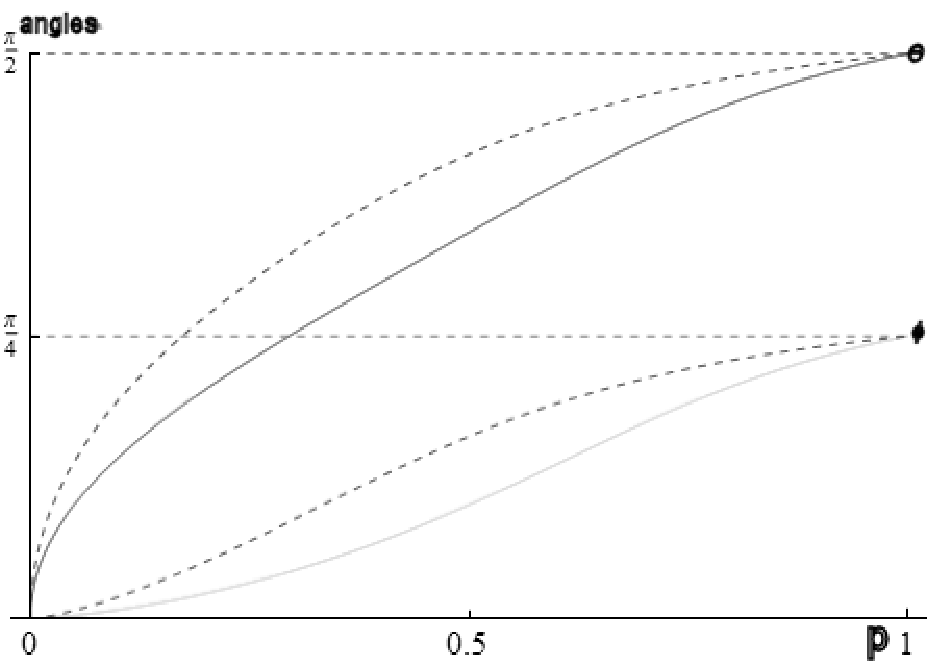, height=6cm} \caption{The
values of the $\theta$ and $\phi$ parameters that correspond to the
maximal Bell operator values (two solid curves concern to the case
$r=1-p$ -- white noise absence; two dashed curves concern to
$r=(1-p)/2$ -- equal weight coefficients for white and colored
noise).}\label{fig:angles_new2}
\end{figure}

In the Fig. \ref{fig:angles_new2} the values of the angles $\theta$
and $\phi$, that provide maximal values of the Bell operator, are
represented. Two solid curves correspond to the case, when in the
twophoton polarization state \eqref{rho_CW} white noise is absent
$(p+r=1)$, and two dashed lines correspond to the case, when colored
and white noise enter into the expression \eqref{rho_CW} with the
same weight $r=(1-p)/2$. Solid curves coincide with the ones plotted
in the work~\cite{Bovino}. From the figure one can see that the
values of the angles $\theta$ and $\phi$ for a fixed pure entangled
state fraction ($p$ is constant) depend on the distribution of
weighting coefficients of white and colored noise. Thus, the
orientation of the analyzers for obtaining maximal values of $\beta$
depends on the fraction distribution between white and colored
noise.

\begin{figure}[tbh!]
\centering \epsfig{file=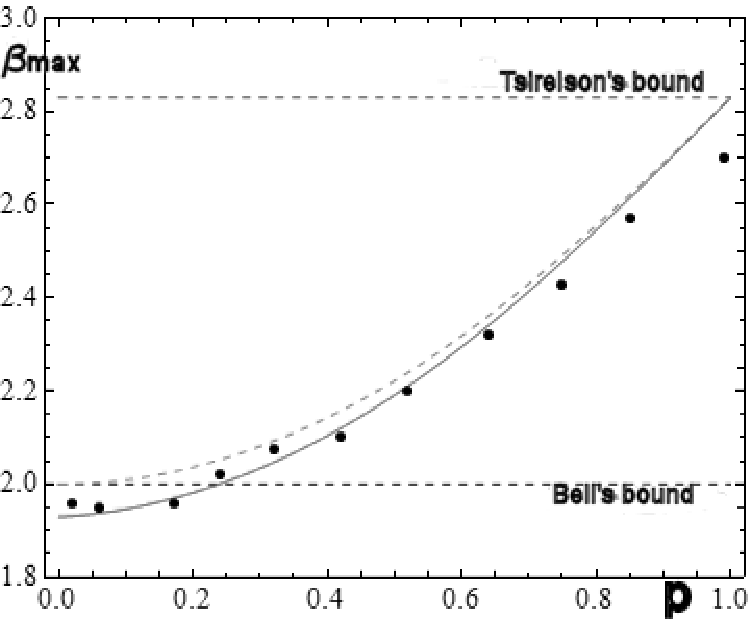, height=7cm} \caption{The points
represent experimental maximal values of $\beta$ from the
work~\cite{Bovino}; the dashed curve is the theoretical prediction
for the maximal values of $\beta$ in the oneparameter colored noise
model~\cite{Cabello}; the solid curve shows theoretical calculations
in the twoparameter (generalized) noise model with the white noise
fraction being $3.5\%$ of the total noise amount in the
system.}\label{fig:exp}
\end{figure}

In the Fig. \ref{fig:exp} the points represent the experimental maximal values of
$\beta$ from the work~\cite{Bovino}; the dashed curve displays
theoretical prediction for the maximal values of $\beta$ on the
oneparameter colored noise model~\cite{Cabello}; the solid curve
illustrates theoretical calculations on the twoparameter
(generalized) noise model with the white noise fraction being
$3.5\%$ of the total noise amount in the system. In the figure we
can see that for such a noise proportion experimental data better
corresponds to theoretical predictions, i.e. the generalized
(twoparameter) noise model is more precise then the oneparameter for
the description of realistic states. But in this case too, as one
can see in the figure, some experimental points lie above and below
the theoretical curve. According to the twoparameter model, this is
explained by the fact that by moving from one point to other not
only does the total noise amount in the system change, but relative
fractions of white and colored noise do too.

\begin{table}[htb!]
\vskip7mm
\centering %\caption{Noise proportions in the system.  Correspondence
%with experimental points in the Fig.\ref{fig:exp}}
%\label{table}
\begin{tabular}{lccccc}
\hline
 Nr. & $p$ & $1-p$ & white,\% & colored,\% & $r$ \\
\hline \hline
 1 & 0.02 & 0.98 & 2 & 98 & 0.96\\
 2 & 0.06 & 0.97 & 3 & 97 & 0.92\\
 3 & 0.17 & 0.83 & 4 & 96 & 0.80\\
 4 & 0.24 & 0.76 & 2 & 98 & 0.75\\
 5 & 0.32 & 0.68 & 2 & 98 & 0.67\\
 6 & 0.42 & 0.58 & 5 & 95 & 0.55\\
 7 & 0.52 & 0.48 & 5 & 95 & 0.46\\
 8 & 0.64 & 0.36 & 7 & 93 & 0.40\\
 9 & 0.75 & 0.25 & 15 & 85 & 0.21\\
10 & 0.85 & 0.15 & 15 & 85 & 0.13\\
\hline \hline
\end{tabular}
\vskip3mm \noindent{\footnotesize{\bf Table of noise proportions in
the system. \\ Correspondence with experimental points in the
Fig.\ref{fig:exp}}} \vskip1mm \noindent\
\end{table}

This kind of interpretation is absolutely logical, because for the
each measurement experimental setup is tuned up in a new way
(particulary, one has to change the analyzers orientation in space).
Remaining in the theoretical model, which is considered in this
work, and choosing the corresponding value of the $r$ parameter
values for each experimental point (for fixed $p$) one can fully
conform theoretical computations with the experimental data. Let us
recall, that the preselected values of the parameters $p$ and $r$,
according to our model, determine the pure entangled state fraction
and relative noise fractions. The percentage of white and colored
noise fractions, that give coincidence between theoretical values
$\beta_{max}$ and experimental data, are represented in the table.
Experimental data were taken from the figure in the
work~\cite{Bovino}.

\section{Conclusions}
For adequate modeling of the twophoton polarization state, created
in the parametric down-conversion process (PDC type II), one should
take into account the presence of colored as well as white noise.

The separability condition for the state $\rho_{CW}$, obtained using
the Perec-Horodecki criterion is the same as the condition obtained
using the majorization criterion. A state $\rho_{CW}$ is separable,
when $3p+r < 1$.

While Bell's inequality violation is extremely robust against
colored noise (Bell's inequality is violated for all $0<p\leq1$),
the violation is unsteady under white noise. White noise presence,
that is determined by a weighting coefficient of just $0.1$
$(p+r=0.9)$, as one can see in the Fig. \ref{fig:contour_new2}, leads to Bell's inequality
violation only for $p\gtrsim0.5$. Simultaneously taking into account
both colored and white noise gives possibility to conform
theoretical computations with experimental data. Taking $p$ and $r$
as adjustable parameters one can determine colored and white noise
fractions by comparison of theoretical calculations with
experimental data.

\end{document}